\documentstyle[preprint,prl,aps]{revtex}
\tightenlines
\begin{document}
\draft
\title{Multipole excitations in quantum dots}
\author{ A. Emperador$^1$, M. Barranco$^2$\cite{perm}, E. Lipparini$^2$,
M. Pi$^1$, and Ll. Serra$^{3}$.}
\address{$^1$Departament ECM,
Facultat de F\'{\i}sica,
Universitat de Barcelona, E-08028 Barcelona, Spain}
\address{$^2$Dipartimento di Fisica, Universit\`a di Trento,
and INFM sezione di Trento, I-38050 Povo, Italy}
\address{$^3$Departament de F\'{\i}sica,
Universitat de les Illes Balears, E-07071 Palma de Mallorca, Spain}
\date{\today}

\maketitle

\begin{abstract}

We have employed  time-dependent local-spin density theory
to analyze the multipole spin and charge density excitations
recently found in GaAs-AlGaAs quatum dots [C. Sch\"uller
et al, Phys. Rev. Lett {\bf 80}, 2673 (1998)]. The overall agreement
between theory and experiment is good, identifying 
the angular momentum of the modes observed in the experiment.
We have found that high multipolarity spin density edge modes
originate from interband transitions instead that from intraband 
transitions, as it happens in the dipole case.

\end{abstract}
\narrowtext
\section*{}

The characteristic single particle and collective excitations of typical
quantum dots (QD) are known to lie in the far-infrared (FIR) energy region,
i.e., they have energies that, depending on the size of the dot, span the range
from a few tens of meV to a fraction of meV.
Experimental information about FIR spectra was first obtained from photon
absorption experiments on InSb and on GaAs quantum dots\cite{Sik89,Dem90}.
Since the confining potential for small dots is parabolic to a good
approximation, and in the FIR regime the dipole approximation works
well, the absorption spectrum is rather insensitive to the number
of electrons in the dot, measuring to a large extent only the
center-of-mass excitations, which at non zero magnetic fields ($B$)
correspond to the two allowed dipole transitions arising from
each of the two possible circular polarizations of the absorbed
light. Two limitations of the absorption process, namely that it is
dominated by the $L$ = 1 multipole of the incoming electromagnetic wave,
and its insensitivity to the electronic spin degree of freedom, have
motivated that  theorists have been mostly concerned with the study of
dipole charge density  excitations (CDE), although
calculations of higher multipolarity density modes 
exist in the literature (see for example Refs.
\onlinecite{Shi91,Gud91,Ye94,Lip97}).

The situation is changing with the use of  inelastic light scattering
to experimentally study QD excitations\cite{Str94,Loc96,Sch96}. In
a way, these studies complement the similar ones carried out in
the past on the two dimensional electron gas (2dEG)\cite{Pin91}.
Besides opening the possibility to study the wave vector dispersion,
inelastic light scattering allows to disentangle charge
from  spin density (SDE) and single particle excitations (SPE), and
to observe them all in the same sample.

Very recently, the $B$ dispersion
of CDE's and SDE's of different multipolarities has been experimentally
determined  in GaAs-AlGaAs quantum dots\cite{Sch98}. We present
here a theoretical interpretation of these results based on the
time-dependent local-spin density theory (TDLSDT)
which addresses for the first time the description of high
multipolarity spin modes.

To this end, we have obtained the ground state (gs) of an $N$ = 200
electron dot  confined  by a uniform, positively
charged disk of $R$ = 120 nm solving the appropiate
Kohn-Sham (KS) equations. These values correspond to a quantum dot
thoroughly studied  by Sch\"uller et al\cite{Sch98}. The
exchange-correlation
energy density ${\cal E}_{xc}(\rho, m)$, where $\rho$ is the electron
density and $m$ the spin magnetization, has been  constructed from
the results of Ref. \onlinecite{Tan89} on the nonpolarized and fully
polarized 2dEG using the two dimensional von Barth and Hedin
prescription\cite{Bar72} to interpolate between both regimes.
The only free parameter in the calculation is the number of positive
charges in the disk, which has been set to $N^+$ = 404 to reproduce
the dipole SDE at $B$ = 0. The range of $B$ values investigated in
this work corresponds to filling factors larger than 3.

Once the KS gs has been worked out, we have determined the induced densities
originated by an external multipole field employing linear-response theory.
Since we  have described at length the  dipole longitudinal
response in dots\cite{Ser98}, we give here only a few details for
presentation purposes. For independent electrons in the KS mean field,
the variation $\delta\rho^{(0)}_{\sigma}$ induced in the spin
density $\rho_{\sigma}$  ($\sigma\equiv\uparrow,\downarrow$) by an external
spin-dependent field $F$, whose non-temporal dependence we denote as
$F=\sum_{\sigma}f_{\sigma}(\vec{r})\,|\sigma\rangle\langle\sigma|$,
can be written as
\begin{equation}
\delta\rho^{(0)}_{\sigma}(\vec{r},\omega) =
\sum_{\sigma'}\int d\vec{r}\,'\chi^{(0)}_{\sigma\sigma'}
(\vec{r},\vec{r}\,';\omega)f_{\sigma'}(\vec{r}\,')\; ,
\label{eq1}
\end{equation}
where
$\chi^{(0)}_{\sigma\sigma'}$ is the KS spin density correlation function
for independent electrons.
In this limit, the frequency $\omega$ corresponds to the
harmonic time dependence of the external field $F$ and  of
the induced $\delta\rho^{(0)}_\sigma$. Eq.\ (\ref{eq1}) is a 2$\times$2
matrix equation in the two-component Pauli space.
In longitudinal response theory, $F$ is diagonal in this space,
and its diagonal components are written as a vector
$F\equiv\left(
\begin{array}{c} f_\uparrow\\ f_\downarrow\end{array}
\right) $.
We consider the external $L$-pole fields
\begin{equation}
F^{(\rho)}_{\pm L}=  r^L e^{ \pm \imath L \theta}
\left(\begin{array}{c} 1\\ 1\end{array}\right)
\,\,\, {\rm and} \,\,\,
F^{(m)}_{\pm L}   =    r^L e^{\pm \imath L \theta}
\left(\begin{array}{c} 1\\ -1\end{array}\right)
\label{eq2}
\end{equation}
which cause, respectively, the charge and spin density $L$-modes.
For the monopole $L$ = 0 mode, these fields are simply taken proportional to
$r^2$. To differentiate the induced densities of each excitation channel
they will be labelled with an additional superscript as
$\delta\rho^{(0,\rho)}_\sigma$ or  $\delta\rho^{(0,m)}_\sigma$

The TDLSDT induced densities are obtained solving the equation
\begin{eqnarray}
\delta\rho^{(A)}_{\sigma}(\vec{r},\omega)
=
\delta\rho^{(0,A)}_{\sigma}(\vec{r},\omega)
+
\sum_{\sigma_1\sigma_2}\int d\vec{r}_1d\vec{r}_2\,
\chi^{(0)}_{\sigma\sigma_1}(\vec{r},\vec{r}_1;\omega)
K_{\sigma_1\sigma_2}(\vec{r}_1,\vec{r}_2)
\delta\rho^{(A)}_{\sigma_2}(\vec{r},\omega) \,\,\, ,
\label{eq3}
\end{eqnarray}
where either $A=\rho$ or $A=m$, and
the kernel $K_{\sigma\sigma'}(\vec{r},\vec{r}\,')$
is the residual two-body interaction.

Equations (\ref{eq3}) have been solved as a generalized matrix
equation in coordinate space. Taking into account
angular decompositions of  $\chi_{\sigma\sigma'}$ and
$K_{\sigma\sigma'}$ of the kind
$K_{\sigma\sigma'}(\vec{r},\vec{r}\,')=
\sum_{\ell}K^{(\ell)}_{\sigma\sigma'}(r,r') e^{i \ell(\theta -\theta')}
$, it is enough to solve this equation for each multipole separately
because only modes with $\ell=\pm L$ couple to the external $L$-pole
field. One has
\begin{eqnarray}
K^{(\ell)}_{\sigma\sigma'}(r,r') &=&
{2\over\pi^{3/2}}
{\Gamma(|\ell|+1/2)\over\Gamma(|\ell|+1)}
{r_<^{|\ell|}\over r_>^{|\ell|+1}}
K_{|\ell|}({r_<\over r_>}) +
\left.{\partial^2{\cal E}_{xc}(\rho,m)
\over\partial\rho_{\sigma}\partial\rho_{\sigma'}}
\right\vert_{gs}{\delta(r-r')\over 2\pi r}\; ,
\label{eq5}
\end{eqnarray}
where $K_n(x)$ is given by the hypergeometric function\cite{Gra80}
${\pi\over2}F(1/2, n+1/2; n+1; x^2)$.

For a polarized system having a non zero magnetization in the gs,
the $\pm L$ modes are not
degenerate and give rise to two excitation branches with
$\Delta L_z=\pm L$, where $L_z$ is the gs orbital
angular momentum.
The induced  charge or magnetization densities corresponding to
density and spin responses are given by
$\delta\rho^{(A)}=\delta\rho^{(A)}_\uparrow+
\delta\rho^{(A)}_\downarrow$ and
$\delta{m}^{(A)}=\delta\rho^{(A)}_\uparrow-
\delta\rho^{(A)}_\downarrow$.
From them, the dynamical polarizabilities in the density and spin channels
are respectively given by
\begin{eqnarray}
\alpha_{\rho\rho}(\ell,\omega)&=&
\int{dr r^{|\ell|+1} \delta\rho^{(\rho)}(r)}\nonumber\\
\alpha_{mm}(\ell,\omega)&=&
\int{dr r^{|\ell|+1} \delta{m}^{(m)}(r)} \; .
\label{eq6}
\end{eqnarray}
For each $L$ value, taking into account both $\pm L$ possibilities we define
$\alpha_{AA}^{(L)}(\omega) \equiv \alpha_{AA}(L,\omega)
+ \alpha_{AA}(-L,\omega)$.
Their imaginary parts are proportional to the
strength functions  $S^{(L)}_{AA}(\omega)=
{\rm Im}[\alpha^{(L)}_{AA}(\omega)]/\pi$.

Figures \ref{fig1} and \ref{fig2} represent the spin and charge strength
functions for $L$ = 0 and 2, respectively. In the quadrupole case we have
indicated with a $+(-)$ sign the SDE's arising from the $+L(-L)$ component
of the $F^{(m)}$ operator in Eq. \ref{eq2} (the low energy CDE is always a
$+$ type excitation, whereas the high energy CDE are $-$ type excitations
arising from the corresponding component of $F^{(\rho)})$.
We have found that the spin peaks are rather fragmented, especially
in the monopole case.  However, they still are collective modes,
with energies redshifted from the free electron ones  due to the attractive
character of the exchange-correlation vertex corrections.

We would like to draw the attention to the  $-$ type, low energy
quadrupole SDE which is seen in Fig. \ref{fig2} to carry an
appreciable strength at $\omega \sim$  3.1 meV and
$B$ = 2 T. When a magnetic field is perpendicularly
applied to a QD, it is well known that the low  energy excitation modes
in the density channel are dipole edge CDE's arising from intraband
transitions while bulk, interband transitions lie higher in energy.
That may change with increasing $L$, and it is particularly
easy to see that this is the case if one looks at the SDE's.
An inspection to  the KS single electron energies shown in Fig. \ref{fig3}
reveals that at high $L$'s, interband electron-hole excitations
are at lower energies than intraband ones.
Since the residual electron-hole interaction is weak in this channel,
we have found that at $B$ = 2 T, the lowest energy octupole
SDE is indeed a mode built from  interband electron-hole excitations.
Still, this is an edge mode, as its existence is only possible
because of the finite size of the system.

Figures \ref{fig4} and \ref{fig5} display the $B$ dispersion of the more
intense SDE's and CDE's, respectively. The solid symbols represent the
experimental data\cite{Sch98}.
It can be seen from these figures that the overall agreement between
theory and experiment is good.
In both spin and charge density
channels, TDLSDT reproduces the weak $B$ dependence of the $L$ = 0 mode
found in the experiment at small $B$ values. Our calculations confirm the
$L$ = 0, 1, and 2  multipolarity assigned  in the experiment to the lower
SDE's, but cannot identify the origin of the higher SDE.
Ruling out the possibility that it is an $L$ = 3 or 4 SDE (see Fig 
\ref{fig4}), it might correspond to a $L <$ 3 mode which gets some 
strength when the dot is probed by an excitation operator carrying a 
finite momentum
$q$ on the dot plane, such as a Bessel function $J_L(qr)$ instead of the
$r^L$ multipole we have been using. Indeed, the signal of that peak is
weak and broad, as mentioned in Ref. \onlinecite{Sch98}.

At $B$ = 0, the energies of the $L > 0$ spin density excitations
follow the simple rule $E_L \sim L E_1$. We attribute this to the weakness
of the residual interaction in the spin channel. The prominent role played
by the strong residual interaction in the charge density channel causes
that rule to fail for CDE's.

As a general trend, the strength carried by the positive $B$ dispersion
branch corresponding to the high $L$ spin density excitations diminishes
as $B$ increases. We have also found that the spin strength
becomes more fragmented with increasing $L$, whereas the bulk and edge
magnetoplasmons associated with the $\pm L$ excitations are  well
defined modes.

The positive $B$ dispersion branches of the  CDE's reveal a
complicated pattern at intermediate $B$ values, quite different from
the expected classical one holding up to $B \sim$ 2-3 T, but that
however fits a large set of the experimental modes.  The behavior of
these branches  has an
interesting quantal origin, namely the formation of well defined Landau
bands for magnetic fields larger than a critical value. Above it, the
more intense high energy collective peaks mostly arise from  transitions
between Landau bands whose index $M$ differs in one unit, $\Delta M=1$.
Since these bands are made of many
single electron states with  different $\ell$ values and energies
rather $\ell$ independent if $B$ is high enough \cite{Pi98},
this explains the
otherwise striking quasi $L$-degeneracy of the plasmon energies, only
broken by finite size effects and the $L$ dependence of the residual
interaction. Other modes with $\Delta M = 2$ build branches satellite
of those formed by the more intense $L$-peaks, which are clearly seen in
the calculation. Satellite branches of this kind appear even in the
dipole case \cite{Dem90,Lor96}, and are a clear signature of
nonparabolic confinement.


This work has been performed under grants PB95-1249 and PB95-0492
from CICYT, Spain, and 1998SGR00011 from
Generalitat of Catalunya. A.E. and M. B. (Ref. PR1997-0174)
acknowledge support from the DGES (Spain).

\begin{figure}
\caption{ Monopole strength function in arbitrary units as a function of energy.
The thick solid line represents the charge density strength, the dashed
line the spin density strength, and the thin solid line the free electron
strength.
}
\label{fig1}
\end{figure}
\begin{figure}
\caption[]{Same as Fig. \ref{fig1} for the quadrupole mode.
}
\label{fig2}
\end{figure}
\begin{figure}
\caption{Single electron energies as a function of orbital angular momentum
$\ell$ for $B$ = 2 T. The horizontal line represents the electron chemical
potential. Full, upright triangles correspond to $\sigma = \uparrow$ states,
and the empty, downright triangles to $\sigma = \downarrow$ states.
Interband and intraband transitions with $\Delta \ell =$ 2, 3 and 4 are
represented to illustrate the energy crossing discussed in the text.
}
\label{fig3}
\end{figure}
\begin{figure}
\caption[]{Energy of the more intense SDE's as a function of $B$. The lines
are drawn to guide the eye, and the solid symbols represent the experimental
data\cite{Sch98}.
}
\label{fig4}
\end{figure}
\begin{figure}
\caption[]{Same as Fig. \ref{fig4} for the more intense CDE's.
}
\label{fig5}
\end{figure}

\begin{references}

\bibitem[*]{perm} Permanent address: Departament d'Estructura i Constituents
de la Mat\`eria, Facultat de F\'{\i}sica, Universitat de Barcelona.
E-08028 Barcelona, Spain.

\bibitem{Sik89} Ch. Sikorski and U. Merkt, Phys. Rev. Lett.
{\bf 62}, 2164 (1989).

\bibitem{Dem90} T. Demel, D. Heitmann, P. Grambow, and K. Ploog,
Phys. Rev.  Lett. {\bf 64}, 788 (1990).

\bibitem{Shi91} V. Shikin, S. Nazin, D. Heitmann, and T. Demel,
Phys. Rev. B {\bf 43}, 11903 (1991).

\bibitem{Gud91} V. Gudmundsson and R. R. Gerhardts, Phys. Rev. B {\bf
43}, 12098 (1991).

\bibitem{Ye94} Z. L. Ye and E. Zaremba, Phys. Rev. B {\bf 50}, 17217 (1994).

\bibitem{Lip97} E. Lipparini, N. Barberan, M. Barranco, M. Pi and Ll.\
Serra, Phys.\ Rev.\ B {\bf 56}, 12375 (1997).

\bibitem{Str94} R. Strenz, U. Bockelmann, F. Hirler, G. Abstreiter, G.
B\"ohm, and G. Weimann,  Phys. Rev. Lett. {\bf 73}, 3022 (1994).

\bibitem{Loc96} D. J. Lockwood, P. Hawrylak, P. D. Wang, C. M. Sotomayor
Torres, A. Pinczuk, and B. S. Dennis,
Phys. Rev.  Lett. {\bf 77}, 354 (1996).

\bibitem{Sch96} C. Sch\"uller, G. Biese, K. Keller,
C. Steinebach, D. Heitmann, P. Grambow, and K. Eberl,  Phys. Rev.
B {\bf 54}, R17304 (1996).

\bibitem{Pin91} A. Pinczuk, D. Heiman, S. Schmitt-Rink, C. Kallin,
B. S. Dennis, L. N. Pfeiffer, and K. W. West, in {\em Light
Scattering in Semiconductor Structures and Superlattices}, pag. 571.
D. J. Lockwood and J. F. Young, editors (Plenum Press, New York, 1991).

\bibitem{Sch98} C. Sch\"uller, K. Keller, G. Biese, E. Ulrichs, L. Rolf,
C. Steinebach, and D. Heitmann,  Phys. Rev. Lett. {\bf 80},
2673 (1998).

\bibitem{Tan89} B. Tanatar and D.M. Ceperley, Phys. Rev. B {\bf 39},
5005 (1989)

\bibitem{Bar72} U. von Barth and L. Hedin,
J. Phys.\ C {\bf 5}, 1629 (1972).

\bibitem{Ser98} Ll. Serra, M. Barranco, A. Emperador, M. Pi, and
E. Lipparini, eprint cond-matt 9806104, to be published in Phys. Rev. B
(1999).

\bibitem{Gra80} I. S. Gradshteyn and I. M. Ryzhik,
{\em Table of Integrals, Series and Products}
(Academic, New York, 1980).

\bibitem{Pi98} M. Pi, M. Barranco, A. Emperador, E. Lipparini,
and Ll.\ Serra, Phys.\ Rev.\ B {\bf 57}, 14783 (1998).

\bibitem{Lor96} A. Lorke, M. Fricke, B. T. Miller, M. Haslinger, J. P.
Kotthaus, G. Medeiros-Ribeiro, and P. M. Petroff, Inst. Phys. Conf. Ser.
{\bf 155}, 803 (1997).

\end{references}
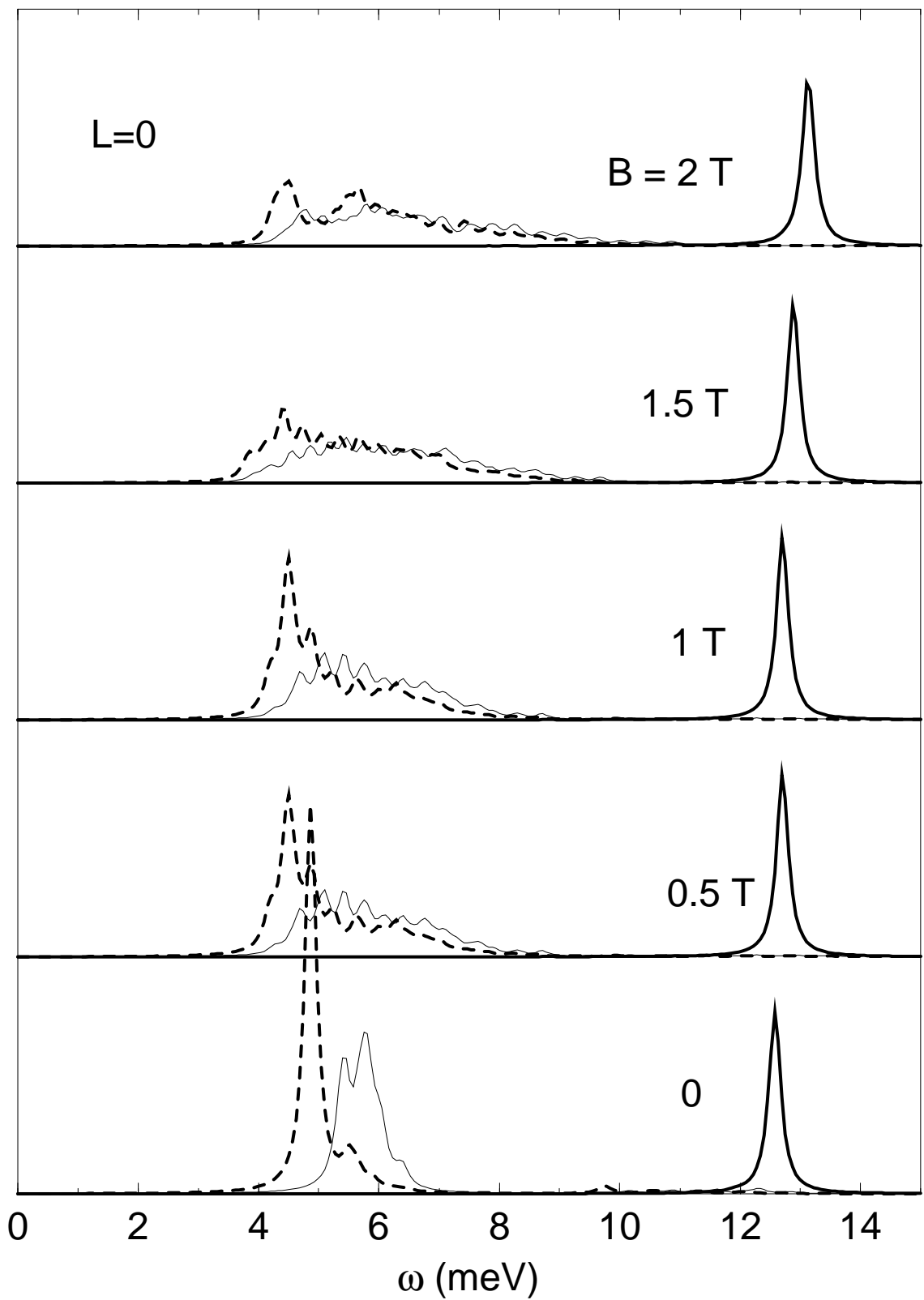
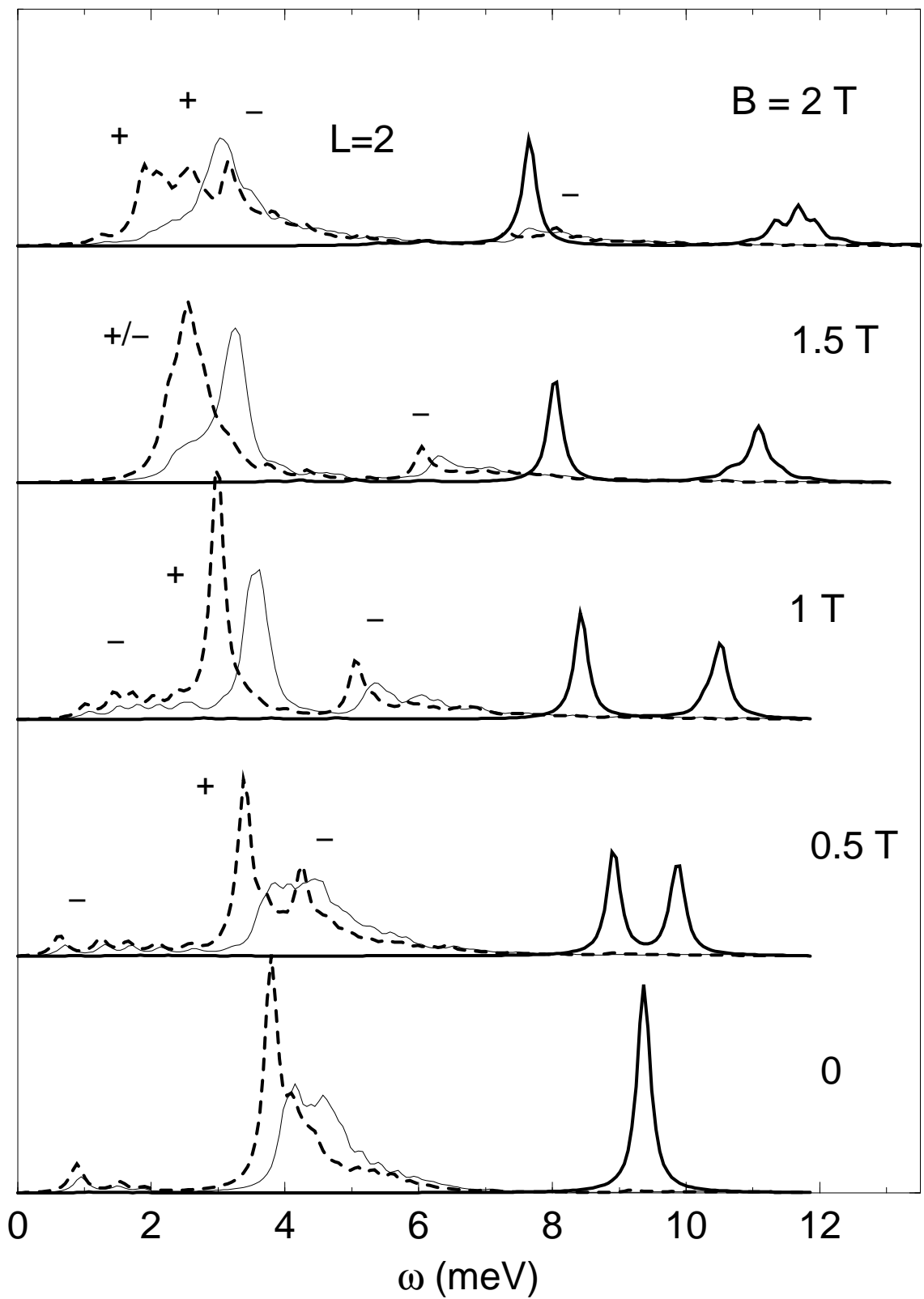
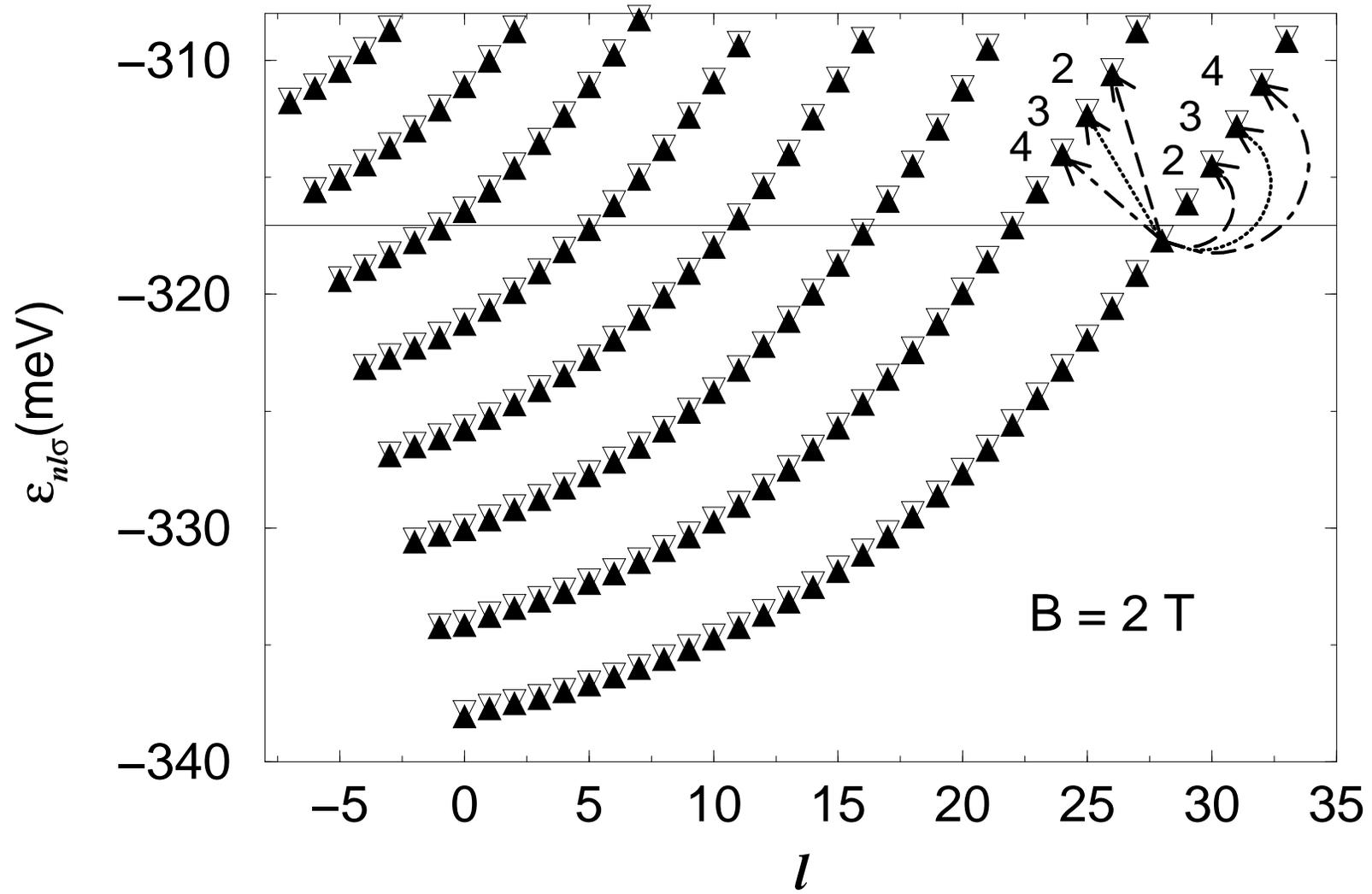
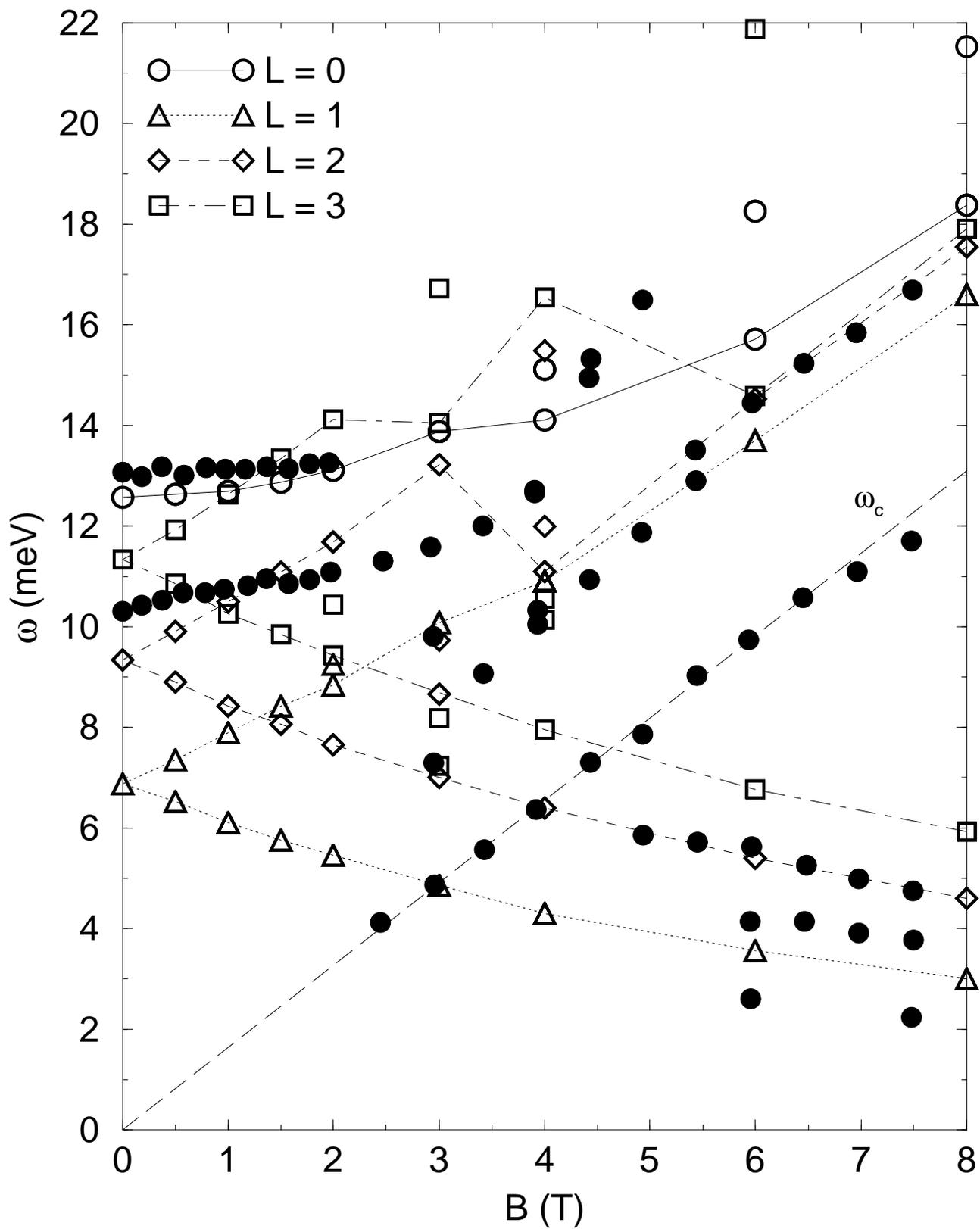
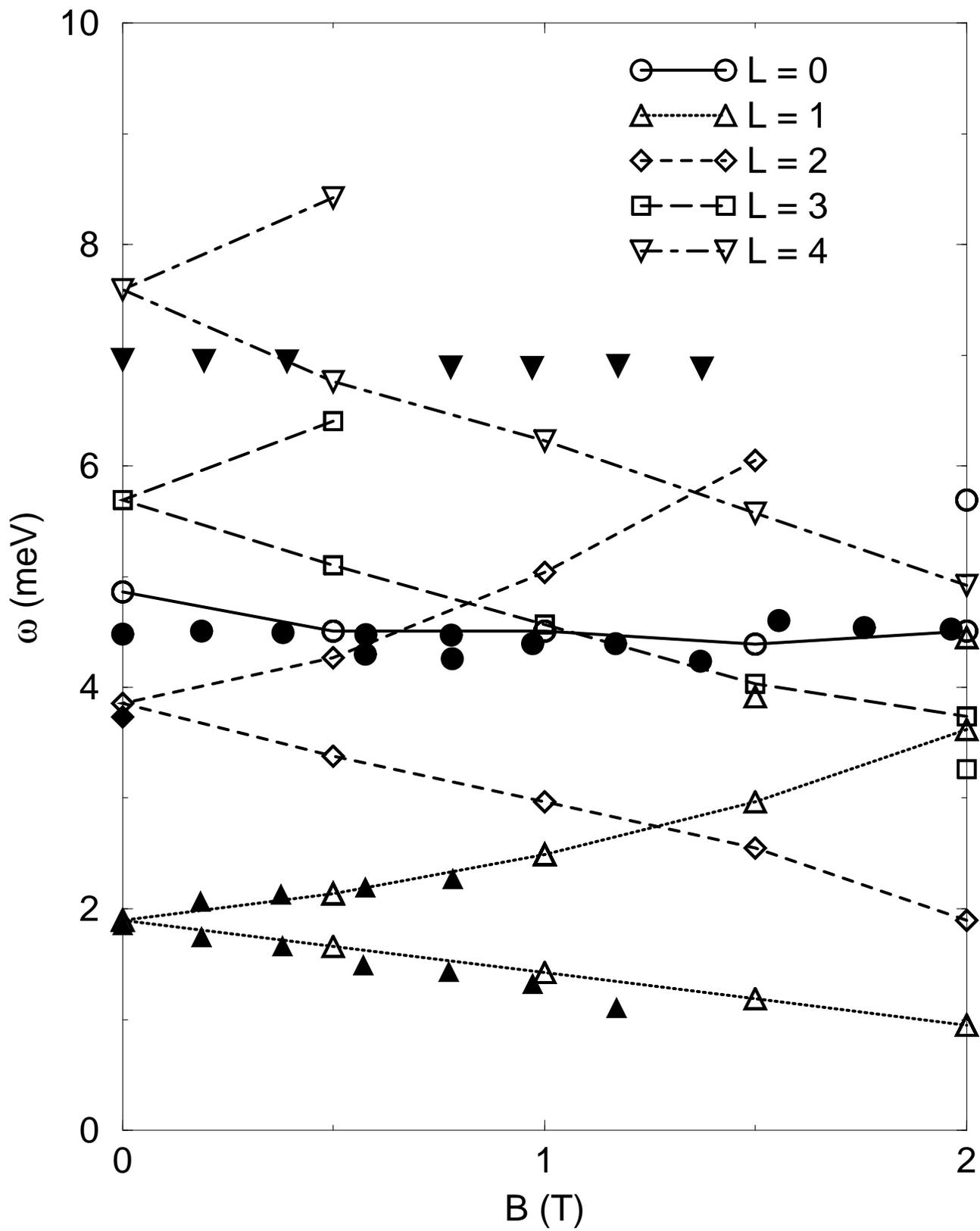
\end{document}